# The REquirements TRacing On target (RETRO).NET Dataset

Jane Huffman Hayes, Jared Payne
Department of Computer Science
University of Kentucky
Lexington, Kentucky, USA

Alex Dekhtyar
Department of Computer Science and
Software Engineering
California State Polytechnic University
San Luis Obispo, California, USA

*Abstract* – **This paper presents the REquirements TRacing On target (RETRO).NET dataset. The dataset includes the requirement specification, the source code files (C# and Visual Basic), the gold standard/answer set for tracing the artifacts to each other, as well as the script used to parse the requirements from the specification (to put in RETRO.NET format). The dataset can be used to support tracing and other tasks.**

## I.  INTRODUCTION

Mission- and safety-critical systems warrant that evidence be provided to show that all safety requirements, security requirements, and functional/non-functional requirements are satisfied by the as implemented software system.  Even software systems that do not have such dire consequences of failure can benefit from knowing the relationships between artifacts of the development lifecycle; such information can support change impact analysis, regression testing, criticality analysis, etc.  *Tracing*, "the activity of either establishing or using traces [1]", is the solution to providing such information/evidence.  In this context, the term *trace* is defined as "the act of following a trace link from a source artifact to a target artifact (primary trace link direction) or vice-versa (reverse trace link direction [1]" and the term *trace link* is defined as "a specified association between a pair of artifacts, one comprising the source artifact and one comprising the target artifact [1]."  Tracing tools generate trace matrices (collections of the links from a source to target artifact).  In order to evaluate the accuracy of a tracing tool, a gold standard or answer set is required (this is often manually generated by domain experts) against which the tracing tool's generated trace matrix is compared.

The traceability community has been very active in developing tracing tools to advance the automated generation of trace links between software engineering artifacts such as requirements, design, code, and test cases [2,3,4,5,6,7,8,12]. To ensure that such tools can scale and can improve the tracing experience for industry practitioners, realistic datasets are needed for tool evaluation.  The lack of datasets for such research is a significant constraining factor.

Datasets are rare for many reasons:  a) the building of a dataset, particularly the determination of the answer set specifying the actual trace relationship between artifacts for the dataset, is not a trivial task; b) real artifacts are rarely available for open source projects; c) companies often considers artifacts to be proprietary and are not willing or able to share them for public consumption; and d) academic researchers receive little or no credit for the time and effort spent building datasets, and consider the time necessary for doing so better spent pursuing other research objectives.

A new approach to developing datasets could be found in the traceability community's very own "backyard" – datasets built based on the tracing tools that the community is developing. Many research groups have such tools:  DeLucia's research group has ADAMS Re-Trace [2], the Hayes and Dekhtyar research group have RETRO [3] and RETRO.NET [9,12], Huang's group has Poirot [4], Niu's group has Tracter [5], Stëghofer research group has Capra [6], Taylor group has ACTS [7], and a group of researchers led by Huang and Poshyvanyk have TraceLab [8].  Each of these tools has a substantial code base, test cases, and sometimes user stories or requirements, design, etc.  A long term goal of the community is to develop datasets for these software projects.  This paper is a first step for this undertaking, a dataset for the RETRO.NET tracing tool [9]. Note that the main goal of the community is to build datasets **for** tracing research, as there is such a dearth.  It just so happens that all of these tracing researchers have tracing tools whose artifacts could be made into tracing datasets.

This paper introduces a dataset based on the RETRO.NET tracing tool [9, 12]. We present a dataset consisting of a set of requirements, the full code base of RETRO.NET, and a traceability matrix (TM), a collection of trace links, tracing each of the requirements to a set of files from the RETRO.NET code base.  The data is presented in RETRO.NET's easy to parse format that can be used by RETRO.NET as well as by some TraceLab components.

The paper is organized as follows:  Section II presents an overview of the dataset, Section III describes how the dataset was developed, and Section IV presents research topics that the dataset can support.  Limitations are listed in Section V.

## II.  OVERVIEW OF DATASET

This section describes the three portions of the dataset, available at DOI 10.5281/zenodo.1223649 and tinycc/seacraft. **Target (Low Level) Artifact Collection: RETRO.NET code base.**   RETRO.NET is a .NET framework- based C# version of RETRO, a.k.a., REquirements TRacing On-Target, a traceability tool built under the direction of the first and the third co-authors (RETRO was the original tool, RETRO.NET is the second generation tool). RETRO originated as a vehicle for implementing a variety of automated requirements tracing techniques developed by our research group [3,13,14]. RETRO functionality was built over the course of three years by a sequence of students working on traceability projects. In 2006, Jody Larsen, then an MS student in our group, observed a number of flaws in the RETRO UI. To facilitate his own research project [12], he elected to build a .NET-based version of the software. Larsen built RETRO.NET from 2006 to 2010, at which point he open-sourced it. Since that time, RETRO.NET has been maintained, incrementally updated, and modified for use in traceability experiments, but has preserved the vast majority of its code base.

RETRO.NET implements the following functionality:
- Creating and maintaining tracing projects
- Tracing of entire dataset at once using a variety of automated requirements tracing methods
- Tracing high-level elements one at a time using automated requirements tracing methods
- Providing a graphical user interface for inspection, approval, and rejection of automatically generated candidate links, and for manual discovery of links
- Incorporating user feedback into the automatic tracing procedure
- Saving the produced TMs in human and machine-readable format (XML)

A screenshot of RETRO.NET is shown on Figure 1. The left side of the RETRO.NET UI contains the source (high-level) artifact broken into individual elements. The selected element has its text shown in the bottom left. The target (low-level) elements that potentially link to the selected source element are shown in the right side of the screen. The text of the selected target element is shown in the bottom right.

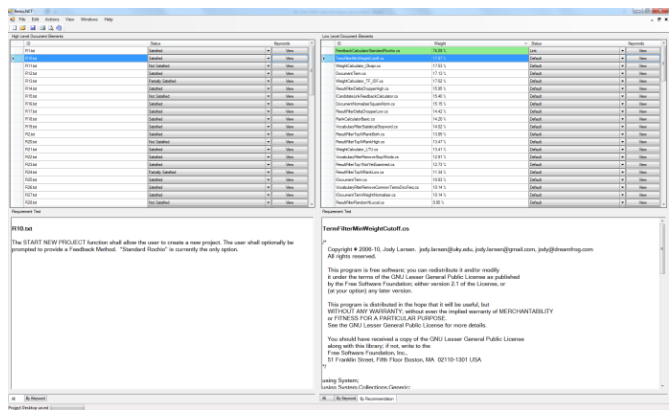

Fig. 1. RETRO.NET screen shot

Despite being eight years old, RETRO.NET is an actively maintained software that is used in a variety of traceability research projects, including a number of studies that took place in 2017 and 2018. RETRO.NET code base released as part of this dataset contains 118 code files. Most of the files contain individual components of the program written in C#. There are several Visual BASIC files, and a couple of project manifest files included for the sake of completeness. The files are released as-is, with native comments accumulated in each file over the course of the software development lifecycle of RETRO.NET. The authors chose not to add any additional comments to the released code base in order not to create bias in future tracing tasks.

**Source (High Level) Artifact Collection: RETRO Requirements.** Our requirements document is a set of 66 functional requirements created by our research team in 2004-2006 as part of the development process for the original RETRO tool [3]. Due to a confluence of circumstances, the development of RETRO.NET did not yield its own requirements document. Note that Larsen, in his work on RETRO.NET, used the RETRO tool, *built to satisfy almost all developed functional requirements from the RETRO requirements specification*, as inspiration both for the tool's functionality and for the improvement of the UI/UX aspects of the tool. Therefore, the RETRO Requirement Specification Version 1.0 that we use in this dataset presents an appropriate, *organically originated* artifact that predates RETRO.NET and contains functional requirements that are relevant for RETRO.NET.

Both the source and the target artifacts are provided in the form that makes them proper inputs for RETRO.NET. Specifically, both source elements and target elements are stored in separate directories (folders), where each file contains a single element. Each requirement is stored as a separate plain text file with the requirement identifier as the file name and the file containing its text. For the sake of completeness, the original requirements specification is also included in the dataset (as a Microsoft Word .docx and a plain text .txt file) along with the python script (parser.py) that was used to parse the text only version of the document subset to create the directory of individual requirements.

**Answer Set (Ground Truth).** The answer set is a collection of links, i.e., pairs consisting of a single source element (an original RETRO requirement) and a single RETRO.NET file. The answer set contains a total of 301 links.

This answer set (a.k.a. that ground truth trace matrix) is provided in XML format as well as a plain text version consisting of source elements followed by tab-separated IDs of all target elements linked to the source element..

### III. DATA PREPARATION PROCESS

Our data collection process had three major steps: preparing the requirements specification, preparing the RETRO.NET code base, and tracing the requirements to the code.

**Preparation of the Requirements Specification.** The preparation of the RETRO Requirements Specification Version 1.0 was discussed in Section II: we took the original requirements document, manually extracted the portion that contained functional requirements, and developed a simple parsing script (parser.py, included with the dataset) to split the functional requirements into individual files placed in a RETRONET-Requirements folder. Additionally, we used the script to capture header level text information and prepend it to individual requirements. For example, if Section 2.1 of the requirements specification simply stated "The ASSESS mode shall permit a user to do N." and a requirement from Section 2.1.1 stated "The user shall be able to exit the ASSESS mode," we placed the following text into the resulting file for the requirement: "The ASSESS mode shall permit a user to do N. The user shall be able to exit the ASSESS mode."

**Preparation of the RETRO.NET code base.** The code files were manually extracted from the Trunk folder of RETRO.NET. First, extraneous folders (such as user manual and system tests, left for future work) were removed from consideration. Then, each folder was "flattened" to form one large folder of files. The files were sorted by type, and each type was examined for inclusion. After manual review of a number of file types, we chose to retain only files with .cs, .vb, and .csproj extensions (other files in the distribution were not code and are not traceable to any requirements). The folder was labeled RETRONET-Trunk.

**Tracing requirements to RETRO.NET files.** We created the traceability matrix provided in this dataset using the following process. Two of the three co-authors, independently of each other, used RETRO.NET to produce two preliminary TMs tracing the RETRO requirements to the RETRO.NET code. After that, they combined the two TMs in a single spreadsheet, and color-coded each link based on whether it was discovered by the first co-author, the third co-author, or both. The co-authors additionally annotated some of the links to formalize and convey their reasoning. On the final stage, the co-authors jointly examined the color-coded list of links and developed the consensus trace. This latter traceability relation is the answer set provided in the dataset.

Because we used the requirements of RETRO, a precursor software system, we were not surprised to see that some of the original RETRO requirements were not satisfied, in fact were abandoned, in RETRO.NET. Some other requirements were traced to some of RETRO.NET files, but were found (through manual inspection) either partially satisfied, or not satisfied, despite the existence of the trace links in the answer set. In essence, we chose to analyze the RETRO.NET feature set by comparing to the original RETRO – it is natural that some early requirements were abandoned in our new tool, and that there are RETRO.NET features not covered by RETRO requirements.

## IV.  RESEARCH TOPICS

The RETRO.NET dataset is a small dataset representative of situations during the software development lifecycle when the development of a formal requirements specification has not caught up with the current state of the software. It can be used to support several different research topics, all falling broadly under traceability: trace link generation, trace matrix assessment, and satisfaction assessment.

Trace link generation research falls into two categories: study of the method and study of the analyst [10]. Under *study of the method*, researchers propose trace link generation methods and/or apply generated trace links from the methods. For this area, the RETRO.NET dataset could be used several ways:
- Propose a new or enhanced trace link generation method and use the dataset to derive recall, precision, f1, f2, and other measures as part of the method's empirical evaluation. A typical research question is: Does method NEW outperform baseline method in terms of recall and precision?
- Apply a new or enhanced trace link generation to a task such as change impact analysis or bug repair. The researchers could apply their method to the dataset, pose a task to be performed, and then derive empirical measures such as those mentioned above plus effort, efficiency, etc. A typical research question is: Does method NEW outperform baseline method in terms of effort and efficiency when performing the NN task?

Under *study of the analyst*, the dataset could be used as follows:
- Propose a new or enhanced trace link generation method and examine the behavior of the analyst as tracing tasks are being performed. A typical research question is: Do analysts using method NEW outperform analysts using baseline method in terms of recall and precision of the final trace matrix?

Under other tasks, the dataset can be used to perform satisfaction assessment and dynamic trace generation. The researchers could use testing or some other technique to determine the actual code files that are executed when various RETRO.NET functions are executed. This could be compared to the static trace generation results shown in the dataset. Further, each requirement could be examined and it could be determined if the code traced to it satisfies the requirement. Automated methods for doing this could be applied, with a typical research question being: Does satisfaction assessment method NEW outperform baseline method or manual method in terms of accuracy and effort?

Finally, researchers who want to use RETRO.NET as a starting point could take the supplied trace matrix and use it to assist them as they make changes to RETRO.NET, to assist them in finding the locations in the code to change. This could be a research study on maintenance of code using automatically generated trace matrices, akin to the study undertaken by Mader et al. [11]. A typical research question is: Can maintenance tasks be performed better using trace matrices generated by method NEW in terms of accuracy and effort as compared to a baseline or manual method?

## V.  LIMITATIONS

There are some limitations to the dataset provided, we address them in terms of threats to validity.

There are threats to external validity. We provide only one dataset that is for one tool in one domain, traceability. We cannot say that this dataset generalizes and could represent all trace generation tools or that it could be used to represent any other domain. However, RETRO.NET has been used by NASA to perform independent verification and validation on real world software systems that are at least mission-critical. Also, RETRO.NET has requirements that are as complex as other real-world systems. The code is as complex as that of other systems, but is poorly commented. Additionally, the dataset only contains two artifacts: requirements and code. We cannot say that it generalizes to represent datasets or projects that may have many other types of artifacts. Also, our artifacts are not very large, there are 66 requirements and 118 code files. There are real world projects that have many more artifacts. Finally, this is not the first dataset released for tracing/traceability research. Rather than considering its limitations in isolation, we ask the reader to view its release as incremental progress towards a *large and diverse* collection of such datasets.

There are threats to internal validity. This mainly is tied to the authors generating the answer set matrix. The authors used RETRO.NET to generate the answer matrix and then manually reconciled the two independent versions to address this threat. During the reconciliation process, the authors established the ground rules for what is considered a link from a requirement to a code file, and have applied these ground rules consistently across all requirements to produce the final trace. Additionally, concerns have been raised that tracing a pair of artifacts describing a requirements tracing tool may somehow be unfair/biased/easier. We do not see it this way. The two artifacts developed organically over a period of time, the authors **did not** have eventual release of these artifacts as a dataset in mind when originally creating them.

Also, the dataset has many artifacts provided in text format. Only one artifact is provided in XML format. Researchers wanting to use the datasets but not with RETRO.NET may have to write their own parsers or data conversion tools.

## ACKNOWLEDGMENT

We thank NSF for partially funding this work under grants CCF-1511117 and CICI 1642134.## REFERENCES